\documentstyle[prd,aps,epsfig]{revtex}
\begin{document}
\title{Lepton asymmetries in exclusive $b \to s\ell^+\ell^-$ 
decays as a test of the Standard Model}
\author{D. Melikhov$^{a}$, N. Nikitin$^{a}$ and S. Simula$^{b}$}
\address{$^{a}$Nuclear Physics Institute, Moscow State University,
Moscow, 119899, Russia\\ $^{b}$ Istituto Nazionale di Fisica Nucleare,
Sezione Sanita, Viale Regina Elena 299, I-00161 Roma, Italy}
\maketitle
\begin{abstract} 
We argue that the longitudinal lepton polarization and the forward-backward
asymmetries in exclusive $B\to (K,K^*)\ell^+\ell^-$ decays are largely
unaffected by the  theoretical uncertainties in the long-distance contributions
both  induced by the long-range charming penguins in the $\ell^+\ell^-$ channel 
and those in the meson channels encoded in the meson transition form factors, 
and are mainly determined by the short-distance Wilson coefficients at the 
low-energy scale $\mu\simeq m_b$. Thus, the above mentioned asymmetries provide 
a powerful probe of the Standard Model and its extensions. 
\end{abstract}
\vspace{.4cm}

The investigation of rare semileptonic decays of the $B$ meson induced 
by the flavour-changing neutral current transition $b\to s$ provides 
an important test of the Standard Model (SM) and its possible extensions. 
Rare decays are forbidden at tree level and occur at the lowest order 
only through one-loop diagrams. This fact opens the
possibility to probe at comparatively low energies the structure of the
electroweak theory at large mass scales, due to the contributions of
virtual particles in the loops. Moreover, rare $b \to s$ transitions are
expected to be sensitive to possible new interactions. 
These interactions govern the structure of the operators and the corresponding 
Wilson coefficients, which appear
in the $\Delta B = 1$ QCD-improved effective Hamiltonian describing the
$b \to s$ transitions at low energies. 

However, any reliable extraction of the perturbative
(short-distance) effects encoded in the Wilson coefficients of the effective
Hamiltonian \cite{inamilim,ali,rescontr,burasmuenz}
requires an accurate separation of the nonperturbative
(long-distance) contributions, which therefore should be known with
high accuracy. The theoretical investigation of these contributions
encounters the problem of describing the hadron structure, which 
provides the main uncertainty in the predictions of exclusive rare
decays. 

In rare exclusive semileptonic decays induced by the quark transition $b\to s
\ell^+\ell^-$  one faces the two types of the long-distance (LD) contributions. 
First, these are the LD effects in the $\ell^+\ell^-$ channel generated by the 
long-range charming
penguins, i.e. intermediate $c\bar c$ states described by the $c\bar c$
resonances ($\psi, \psi'$, ...)  and the charmed hadronic continuum states. 
And second, these are the LD effects in the meson transition amplitude of the
Effective Hamiltonian encoded in the meson transition form factors. 

In this letter we study the influence of the above mentioned two types of the LD
effects on observables in exclusive $B\to K^{(*)}\ell^+\ell^-$ decays. We show
that the asymmetries of the  lepton distributions remain largely unaffected by
the uncertainties in the LD effects but  turn out sensitive to the values of the
Wilson coefficients thus providing a possibility to reliably separate and study
the short-distance (SD) effects. 

\noindent 1. Neglecting the strange quark mass the effective Hamiltonian
describing the $b\to s\ell^+\ell^-$ transition has the following structure
\cite{ali,burasmuenz} 
\begin{eqnarray}
\label{heff}
{\cal H}_{eff}(b\to 
sl^{+}l^{-})&=&{\frac{G_{F}}{\sqrt2}}{\frac{\alpha_{em}}{2\pi}}
\lambda_t                                            
\left[\,-2i{\frac{C_{7\gamma}(\mu)}{q^2}}
m_b{\bar s_L}\sigma_{\mu\nu}q^{\nu}b\cdot{\bar l}\gamma^{\mu}l
\right.\\&&+\left. 
C_{9V}(\mu)\bar s_L \gamma_\mu b\cdot{\bar l}\gamma^{\mu}l+
C_{10A}(\mu){\bar s_L} \gamma_{\mu}b\cdot{\bar 
l}\gamma^{\mu}\gamma_{5}l\right],\nonumber
\end{eqnarray}
where $\lambda_t=V^{*}_{ts}\,V_{tb}$ and $\bar s_L\equiv \bar s(1+\gamma_{5})$. 
The quantities $C_i(\mu)$ are the SD Wilson 
coefficients which are obtained by integrating out the  heavy particles. 
The values of the Wilson coefficients
at the scale $\mu\simeq 5\;GeV$ are \cite{burasmuenz}: 
$C_1(\mu) = 0.241$,
$C_2(\mu) = -1.1$, 
$C_{7\gamma}(\mu) = 0.312$, 
$C_{9V}(\mu) =-4.21$ and 
$C_{10A}(\mu)= 4.64$. 

An important contribution to the $b\to s\ell^+\ell^-$ transition is given by the 
intermediate $c\bar c $ states in the $\ell^+\ell^-$ channel, generated by the 
4-quark
operators in the effective Hamiltonian describing the $b\to s$ transition, and 
proceeding 
through the chain of transitions $b\to sc\bar c\to s\gamma^*\to s\ell^+\ell^-$. 
The $c\bar c$ pairs in this channel generate both the SD and LD contributions. 

The $b\to s\gamma^*$ transition vertex is transverse with respect to the photon 
momentum because
of the gauge invariance. In the limit $m_s=0$ within a theory with a $V-A$ 
charged current
it has the general structure of the form 
\begin{eqnarray}
\label{bsgamma}
\langle s(k_2)|J_{\mu}^{em}(0)|b(k_1) \rangle =
\frac23|e|\frac{G_F}{\sqrt2}\lambda_t(C_1+C_2/3) 
\bar s_L(k_2)\left[(\gamma_\mu q^2-q_\mu \hat q)G_1(q^2)
-2im_b \sigma_{\mu\nu}q^\nu G_2(q^2)\right]b(k_1),
\end{eqnarray}
where $G_1$ and $G_2$ are the independent form factors. 
It is important to point out that the form factor $G_1(q^2)$ has no pole at 
$q^2=0$. This property is related to the fact that the photon mass remains 
unrenormalized at all orders in perturbation theory. Also notice that the 
pole in $G_1(q^2)$ would have yielded a nonzero amplitude of the radiative 
$B\to K\gamma$ decay. 

The amplitude of the $b\to s\ell^+\ell^-$ transition through the $c\bar c$ 
penguin then has the form 
\begin{eqnarray}
A(b\to sc\bar c\to s\gamma^*\to s\ell^+\ell^-)=
-\frac{G_F}{\sqrt2}\lambda_t (C_1+C_2/3) \frac{2}{3}e^2
\bar s_L\left[\gamma_\mu q^2 G_1^{c\bar c}(q^2)-
2im_b \sigma_{\mu\nu}q^\nu G_2^{c\bar c}(q^2)\right]b\cdot
\bar l\gamma_\mu l,
\end{eqnarray}
where $G^{c\bar c}_{1,2}$ are the $c\bar c$ contributions to $G_{1,2}$. 

The Lorentz structure of this amplitude is similar to the structure of the effective 
Hamiltonian (\ref{heff}),   
so the $c\bar c$ contributions to the $b\to  s\ell^+\ell^-$ transition can be 
translated into 
additions to the Wilson coefficients as follows 
\begin{eqnarray}
C_{9V}&\to &C_{9V}^{eff}(q^2)=C_{9V}+\Delta C^{c\bar c}_{9V}(q^2),\qquad 
\Delta C^{c\bar c}_{9V}(q^2)\equiv \frac{16\pi^2}{3}(C_1+C_2/3)G^{c\bar 
c}_1(q^2)\\
C_{7\gamma}&\to &C_{7\gamma}^{eff}(q^2)=C_{7\gamma}+\Delta C^{c\bar 
c}_{7\gamma}(q^2),\qquad
\Delta C^{c\bar c}_{7\gamma}(q^2)\equiv \frac{16\pi^2}{3}(C_1+C_2/3)G^{c\bar 
c}_2(q^2). 
\nonumber
\end{eqnarray}

In the $q^2$-region of the vector $c\bar c$ resonances ($V$) the form factors 
$G^{c\bar c}_{1,2}$ have 
singularities corresponding to these resonances. The general expression for the 
$b\to sV$ 
amplitude can be written as (cf. \cite{soares}) 
\begin{eqnarray}
A(b\to sV)=
-\frac{G_F}{\sqrt2}\lambda_t(C_1+C_2/3)
\bar s_L\left[\gamma_\mu M_V^2g_1(M_V^2)-2im_b \sigma_{\mu\nu}q^\nu 
g_2(M_V^2)\right]b
\epsilon^V_\mu(q),
\end{eqnarray}
where $\epsilon^V_\mu(q)$ is the $V$ polarization vector.  
Then the form factors $G^{c\bar c}_{1,2}(q^2)$ in the region $q^2\simeq M_V$ 
are expressed through the $g_{1,2}$ as follows 
\begin{eqnarray}
\label{G}
G_i^{c\bar c}(q^2)=\frac{g_i(M_V^2)}{M_V^2-q^2-i\Gamma_V M_V}f_V M_V+
[{\rm regular\; terms\;at\;} q^2=M_V^2].
\end{eqnarray}
The leptonic decay constant $f_V$ is defined by the relation  
$\langle 0|\bar c\gamma_\mu c|V\rangle=\epsilon_\mu^V(q)M_V f_V$.  

If we neglect the soft nonfactorizable gluon exchanges, the amplitude of the 
$b\to sl^+l^-$ transition
through the charming penguin is connected with the charm contribution to
the vacuum polarization, $\Pi^{c\bar c}_{\mu\nu}$, defined as 
\begin{equation}
\Pi^{c\bar c}_{\mu\nu}(q)=\frac{1}{i}\int dx e^{-iqx}\langle 0|T(\bar 
c(0)\gamma_\mu c(0), \bar
c(x)\gamma_\nu c(x))|0\rangle. 
\end{equation}
As a consequence of the vector current conservation, $\Pi^{c\bar c}_{\mu\nu}$ 
has the transverse structure
\begin{eqnarray}
\Pi^{c\bar c}_{\mu\nu}=(g_{\mu\nu}q^2-q_\mu q_\nu)\Pi^{c\bar c}(q^2).  
\end{eqnarray}
The quantity $\Pi^{c\bar c}(q^2)$ contains both the LD and SD contributions and 
does not have 
a pole at $q^2=0$ because of the gauge invariance. 
In this factorization approximation one has 
\begin{eqnarray}
G_1^{c\bar c}=\Pi^{c\bar c}(q^2),\qquad G_2^{c\bar c}=0. 
\end{eqnarray}
Similarly, assuming factorization for the $b\to s V$ amplitude, one finds
\begin{eqnarray}
g_1(M_V^2)=f_V/M_V,\qquad g_2(M_V^2)=0, 
\end{eqnarray}
and relation (\ref{G}) takes a simple form 
\begin{eqnarray}
\label{G1}
G^{c\bar c}_1(q^2)=\frac{f_V^2}{M_V^2-q^2-i\Gamma_V M_V}
+[{\rm regular\;terms\;at\; } q^2=M_V^2], \qquad G^{c\bar c}_2(q^2)=0, 
\end{eqnarray}
where the regular terms come from the contribution of other resonances and 
the $c\bar c$ hadronic continuum states. A standard procedure 
\cite{ali,rescontr}
relies in assuming the $c\bar c$ 
hadronic continuum to be described by the $c-$quark loop in the spirit of the 
quark-hadron duality. 
Then, summing over all relevant resonances and re-expressing the decay constants
through the corresponding  leptonic decay rates with the relation 
$$\Gamma(V\to l^+l^-)=\pi\alpha_{em}^2\frac{16}{27}\frac {f_V^2}{M_V}$$ 
we come to the following representation for the shift of the Wilson coefficient 
$C_{9V}$ in the 
factorization approximation (cf.\cite{ali,rescontr})
\begin{equation} 
\label{c9}
\Delta C^{c\bar c}_{9V}(q^2)=[3C_1(m_b)+C_2(m_b)]\cdot
\left[h({m_c/m_b}, {q^2/m_b^2}) +
\frac{3}{\alpha_{em}^2}\kappa\sum_{V=J/\psi,\psi',\cdots}
\frac{\pi \Gamma(V \to \ell \ell) M_{V}}{M_{V}^2 - q^2 - i
M_{V} \Gamma_{V}} \right]. 
\end{equation}  
The fudge factor $\kappa$ is introduced in Eq. (\ref{c9}) to account for 
inadequacies of the
naive factorization framework (see \cite{neubertstech} for more details). 
Phenomenological analyses \cite{rescontr} suggest that in order to reproduce 
correctly the branching ratio 
${\rm BR}(B\to J/\psi X\to\ell^+\ell^-X)=$ ${\rm BR}(B\to J/\psi X)~\cdot$ 
${\rm BR}(J/\psi\to\ell^+\ell^-)$ it should satisfy an approximate 
relation $\kappa ~ \left[ 3 C_1(m_b) + C_2(m_b) \right] \approx -1$. 
The SD contributions are contained in the function $h(m_c/m_b,q^2/m_b^2)$, 
which describes the one-loop matrix element of the four-quark operators (see, e.g., 
\cite{burasmuenz} for its explicit expression). 

It is known that the charm contribution described by (\ref{c9}) yields a strong 
interference between the LD
and SD contributions to the decay amplitude in a broad range of $q^2$. It has 
been argued however that 
a simple parametrization of the hadronic $c\bar c$ continuum by the quark loop 
in the spirit of the quark-hadron
duality considerably overestimates the net effect of the charm at small $q^2$. 
One possibility to describe
the $c\bar c$ continuum contribution at small $q^2$ in a more realistic way is 
to take into account the
regular terms in (\ref{G}) by representing the LD contributions to the $C_{9V}$ 
as a sum of the
Breit-Wigner resonances but assuming a $q^2$-dependent $f_\psi(q^2)$ 
\cite{ahmady}. 
Another possibility has been implemented in Ref. \cite{kruger} where the $c\bar 
c$ contribution to $C_{9V}$
has been described as a sum of the Breit-Wigner resonances and the continuum 
contribution 
which has been expressed through the observable cross-section of the process 
$\ell^+\ell^-\to$ charmed hadrons. 

Since at present there are no firm arguments to determine with a sufficient 
accuracy the LD $c\bar c$ effects 
we consider the difference provided by the three models of the $\Delta C^{c\bar 
c}_{9V}$, 
namely, the one given by Eq. (\ref{c9}), and 
those of Refs. \cite{ahmady,kruger}, as a typical present-day theoretical
uncertainty in $C_{9V}$.  The corresponding $C_{9V}^{eff}(q^2)$ are 
plotted in Fig. 1. 
We point out once more that the pole at $q^2=0$ in $C_{9V}^{eff}$ which has been 
discussed in \cite{luzhang} is
ruled out by the gauge invariance. 

Let us now briefly discuss the nonfactorizable effects. First, notice that 
factorization yields vanishing 
of the $c\bar c$ contribution to the amplitude of the radiative $b\to s\gamma$ 
decay at $q^2=0$: 
in this case $G_2=0$ and the structure proportional to 
$G_1$ does not contribute at $q^2=0$. Hence, the LD $c\bar c$ contribution to 
the radiative $b\to s\gamma$ 
transition is a purely nonfactorizable effect. It should be also taken into 
account that a nontrivial $c\bar c$ 
contributions to the radiative $b\to s\gamma$ decay $\Delta C^{c\bar 
c}_{7\gamma}(0)\ne 0$ 
implies a nonzero $\Delta C^{c\bar c}_{7\gamma}(q^2)$ in the 
semileptonic $b\to s\ell^+\ell^-$ decay. 

A stringent test of the factorization or its breakdown is  
provided for instance by the polarization of $\psi$ in the inclusive and 
exclusive $b\to s\psi$ 
transitions: the ratio of the longitudinally-to-transversely polarized $\psi$ 
does not depend on 
the Wilson coefficients but is rather a function of $g_1(M^2_\psi)$ and
$g_2(M^2_\psi)$.  The analysis of Ref. \cite{soares}  favours the value 
$|g_2(M^2_\psi)/g_1(M^2_\psi)|\simeq 0.1\div 0.2$, although the
factorization prediction $g_2(M^2_\psi)=0$ is also not ruled out. 
For estimating at all kinematically accessible $q^2$ the nonfactorizable effects 
which lead to a nonzero 
$\Delta C^{c\bar c}_{7\gamma}$ it seems reasonable to assume that the ratio 
$G_2/G_1$,
which governs the size of these effects, is limited by the relation 
$|G_2(q^2)/G_1(q^2)|\le 0.2$ 
in the whole kinematically accessible region of $q^2$. In this case, the size of 
the nonfactorizable effects in the
region of the resonances corresponds to the analysis of \cite{soares}, and at 
$q^2=0$ this prescription   
yields a nonfactorizable contribution to the $b\to s\gamma$ decay within $5\%$ 
in accordance with the
estimates of Ref. \cite{nonfact}. 
In the following analysis of the sensitivity of the asymmetries in exclusive 
$B\to K^{(*)}\ell^+\ell^-$ 
decay to the uncertainties in the LD effects we also allow for a nonfactorizable 
contribution 
$|\Delta C^{c\bar c}_{7\gamma}(q^2)|\le 0.2 |\Delta C^{c\bar c}_{9V}(q^2)|$. 

\noindent 2. The amplitude of the $B\to K^{(*)}\ell^+\ell^-$ decay is given by 
the relevant mesonic matrix element of
the effective Hamiltonian. The LD effects connected with the meson formation  in
the initial and final $q\bar q$ channels are encoded in the meson transition
form factors of the bilinear quark currents from the  effective Hamiltonian
(\ref{heff}).  Various theoretical frameworks have been
applied to the description of meson transition form factors: among them are
constituent quark models \cite{jauswyler,stech,mns}, 
QCD sum rules \cite{colangelo,aliev,damir}, lattice QCD \cite{lat1,lat}, 
analytical constraints \cite{lellouch}. 

Lattice QCD simulations, because of its most direct connection with QCD, 
are expected to provide the most reliable results. 
However, at present the lattice calculations do not provide the form factors 
in the whole accessible kinematical decay region as  the daughter light quark 
produced in $b$ decay cannot move fast enough on the lattice
and one is therefore limited to the region of not very large recoils. 
For obtaining form factors in the whole kinematical decay region one can 
use extrapolation procedures based on some parametrizations of the form
factors. For instance, in \cite{lat} a simple lattice-constrained 
parametrization based 
on the constituent quark picture \cite{stech} and pole dominance is developed. 

QCD sum rules give complementary information on the form factors 
as they can calculate the latter at not very large momentum 
transfers. However in practice various versions of the QCD sum 
rules give remarkably different predictions, being strongly
dependent on the technical subtleties of the particular version 
\cite{colangelo,aliev,damir}.  

Constituent quark models (QM) have proved to be a fruitful 
phenomenological method for the description of heavy meson transitions. 
A long-standing shortcoming of the quark model predictions for the form factors 
has been a strong dependence of the results on the QM parameters \cite{mns}. 
However, as we have found recently \cite{mns1}, a combination of the results of  
the lattice simulations at large $q^2$ and the dispersion approach based on 
the constituent quark picture \cite{m} 
allows one to considerably increase the 
accuracy of the predictions. Namely, constraining the parameters of the quark 
model by the 
requirement that the form factors of the QM at small recoils reproduce the 
results of the lattice 
simulations considerably decreases the uncertainty in the relevant QM
parameters.  Once these parameters are fixed, the spectral representations of
our dispersion QM   allows a direct calculation of the form factors in the whole
kinematically accessible region  of $q^2$. The form factors of the dispersion QM
develop  the correct heavy-quark expansion at leading and next-to-leading
$1/m_Q$ orders  in accordance with QCD for the transitions between heavy quarks;
for the heavy-to-light transition  the form factors of the dispersion QM satisfy
the relations between  the form factors of vector, axial-vector, and tensor
currents valid at small recoil.  In addition they satisfy the dispersive bounds
and thus, as well as the form factor parametrizations of Ref.  \cite{lat},
satisfy all known rigorous theoretical constraints.  However some uncertainties
are still present which are connected with the errors in the results of the 
lattice simulations at small recoil and the approximate character of the
extrapolating formulas  used in \cite{lat} as well as the constituent quark
picture used for the description of the form factors  in Ref. \cite{mns1}. We
consider that the difference between the form factors  of Refs. \cite{lat,mns1}
provide a typical theoretical uncertainty of our understanding of the LD
effects  encoded in the transition form factors. 

\noindent 3. We evaluate the forward-backward ($A_{FB}$) and the longitudinal 
lepton polarization ($P_L$) asymmetries 
for various parametrizations of the LD $c\bar c$ contributions to the 
Wislon coefficients $C_{9V}^{eff}$ and $C_{7\gamma}$ and various sets of the 
form factors with formulas given in \cite{mns1}. 
Fig. 2 shows these quantities evaluated with the GI-OGE set of
the form factors from \cite{mns1} and $\Delta C_{7\gamma}^{c\bar c}=0$ 
and various prescriptions of $C_{9V}^{eff}(q^2)$. 
Fig. 3 demonstrates a sensitivity of the asymmetries to the
nonfactorizable effects parametrized as additions to the Wilson coefficient
$C_{7\gamma}$. One can see that the uncertainties due to the LD $c\bar c$
effects in both $C_{7\gamma}$ and $C_{9V}$ provide a minor influence on the 
observable quantities in a broad kinematical region beyond the resonances. 
Notice that within the SM 
$P_L(B\to K\mu^+\mu^-)\simeq 2C_{9V}C_{10A}/(C^2_{9V}+C^2_{10A})\simeq -1$ 
independently of the prescriptions chosen for the LD effects in the Wilson 
coefficients. 

Fig. 4 presents the same asymmetries evaluated with the $C_{9V}$ given by the 
Eq. (\ref{c9}) and 
$C^{c\bar c}_{7\gamma}=0$ 
and the two sets of the form factors: namely, the lattice-constrained 
parametrization of Ref. \cite{lat} 
and the GI-OGE set from Ref. \cite{mns1}. One can observe that the asymmetries 
turn out to be more sensitive
to the variations of the transition form factors with respect to the
uncertainties in the LD effects induced by the $c\bar c$ penguins. 
Nevertheless, there is still a broad range of momentum transfers in which the
interference between the LD and SD contributions is negligible. Thus, a study
of the asymmetries under consideration in these regions provides a potential
possibility to measure the Wilson coefficients. 

In fact, a sensitivity of the asymmetries to the SD contributions encoded in 
the Wilson coefficients might be observed. The CLEO data on the radiative 
exclusive and inclusive $b\to s\gamma$ decays allow 
$R_{7\gamma}=C_{7\gamma}(M_W)/C^{SM}_{7\gamma}(M_W)$ 
in the ranges (see, e.g. \cite{misiak}) 
$0.4\le R_{7\gamma}\le 1.2$, and $-4.2\le R_{7\gamma}\le -2.4$. 
As discussed in Refs. \cite{misiak,hewett} scanning over the MSSM parameter 
space correspondent to the allowed
$C_{7\gamma}$ regions provides only minor changes in $C_{9V}(M_W)$ and 
$C_{10A}(M_W)$. 
Thus for illustrating possible new physics effects in lepton 
asymmetries we fix $C_{9V}(M_W)$ and $C_{10A}(M_W)$ to their SM values, 
and vary $C_{7\gamma}$ in the allowed regions. 
The corresponding $A_{FB}$ and $P_L$ are shown in Fig. 5. 
One finds the shape of $A_{FB}$ to be sensitive to the $C_{7\gamma}$ value 
(or at least to the $C_{7\gamma}$ sign), and this effect far 
overwhelms the uncertainties in the LD contributions. 

A weaker but still visible sensitivity of $P_L(B\to K^*\ell^+\ell^-)$ to 
the SD coefficients can be seen (Fig. 5. b). 

The longitudinal lepton polarization asymmetry $P_L(B\to K\mu^+\mu^-)$ is 
equal to $P_L\simeq 2C_{9V}C_{10A}/(C^2_{9V}+C^2_{10A})$ if $C_{7\gamma}$ lies 
in the allowed regions at all kinematically accessible $q^2$, except for 
the end-points and regions near $\psi$ and $\psi'$. 
Therefore, the measurement of $P_L(B \to K \mu^+ \mu^-)$ can provide us 
direct information on the ratio $C_{9V}/C_{10A}$.  

In conclusion, we have analyzed the sensitivity of the lepton asymmetries in 
rare
exclusive $b\to s\ell^+\ell^-$ transitions to the LD effects. We have found the
most important uncertainty to come from the errors in the  meson transition form
factors whereas the uncertainties in the LD effects induced by the long-range
$c\bar c$  penguins are of relatively minor importance. So for obtaining better
predictions more accurate  calculations of the transition form factors in the
whole accessible $q^2$ region are necessary.  Nevertheless, there are broad
$q^2$-intervals where the interference between the LD and SD effects is
negligible and the asymmetries in these regions contain information on the SD
effects. 

In particular, the shape of the forward-backward asymmetry in $B\to 
K^*\mu^+\mu^-$ turns out to be specific 
for the particular values of the SD Wilson coefficients: namely, within the SM, 
$A_{FB}$ is positive at 
small $q^2$, has a zero at $q^2\simeq 0.15 M^2_B$ and then becomes negative at 
larger $q^2$. At the same 
time, the present experimental restrictions allow a region of the $C_{7\gamma}$ 
with a sign opposite to 
the SM value which yields an essentially different shape of the $A_{FB}$ (cf. 
also \cite{burdman}). 
The longitudinal lepton polarization asymmetry $P_L(B\to K^*\mu^+\mu^-)$ also 
contains important
information on the Wilson coefficients at the scale $\mu\simeq m_b$, and 
$P_L(B\to K\mu^+\mu^-)$ 
directly measures $C_{9V}/C_{10A}$ at $\mu\simeq m_b$. 

Thus, the experimental study of the forward-backward asymmetry and the
longitudinal lepton polarization  asymmetry potentially provides an effective
test of the Standard Model and its possible extensions. 

We would like to thank M. Misiak for helpful comment on the MSSM parameter 
space.

\begin{center}
\begin{figure}
\mbox{\epsfig{file=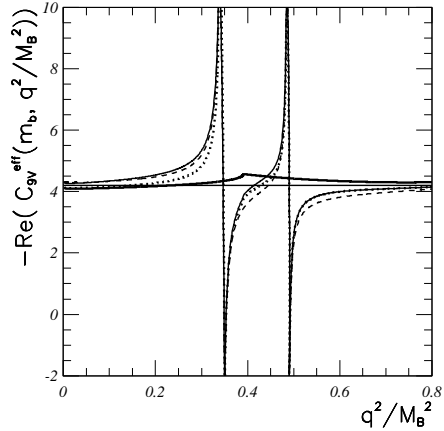,width=7.cm}}
\caption{\label{fig:c9}
The effective Wilson coefficients $C_{9V}^{eff}(q^2)$ for various 
parametrizations of the $c\bar c$ 
contributions: solid - Eq. (\protect\ref{c9}), dotted - \protect\cite{ahmady}, 
dashed - \protect\cite{kruger}. 
The bold solid line corresponds to $C_{9V}$ without the LD contributions, 
namely, 
$C_{9V}(\mu)+(3C_1+C_2)h(m_c/m_b,q^2/m_b^2)$, the horizontal line corresponds to 
$-C_{9V}(\mu=5\;Gev)=4.21$.}
\end{figure}
\end{center}

\newpage

\begin{center}
\begin{figure}
\begin{tabular}{cc}
\mbox{\epsfig{file=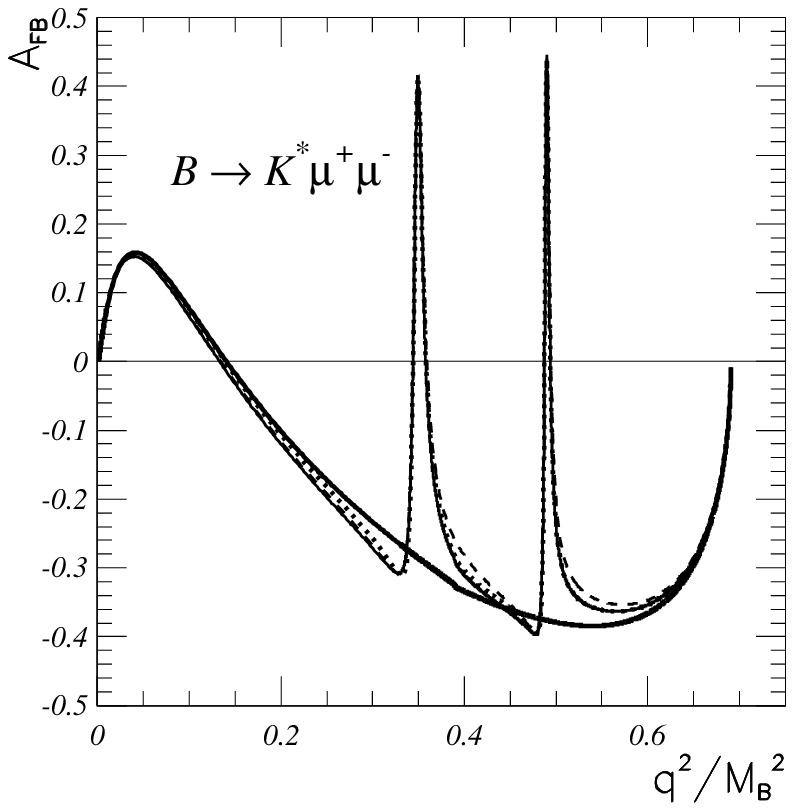,width=7.5cm}}
& 
\mbox{\epsfig{file=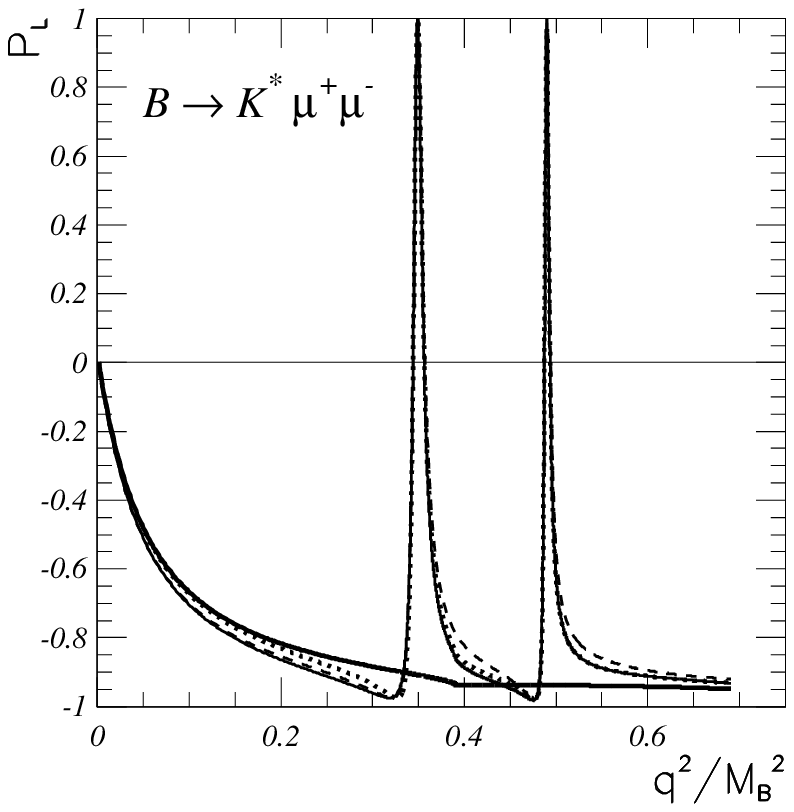,width=7.5cm}}
\end{tabular}

\mbox{\epsfig{file=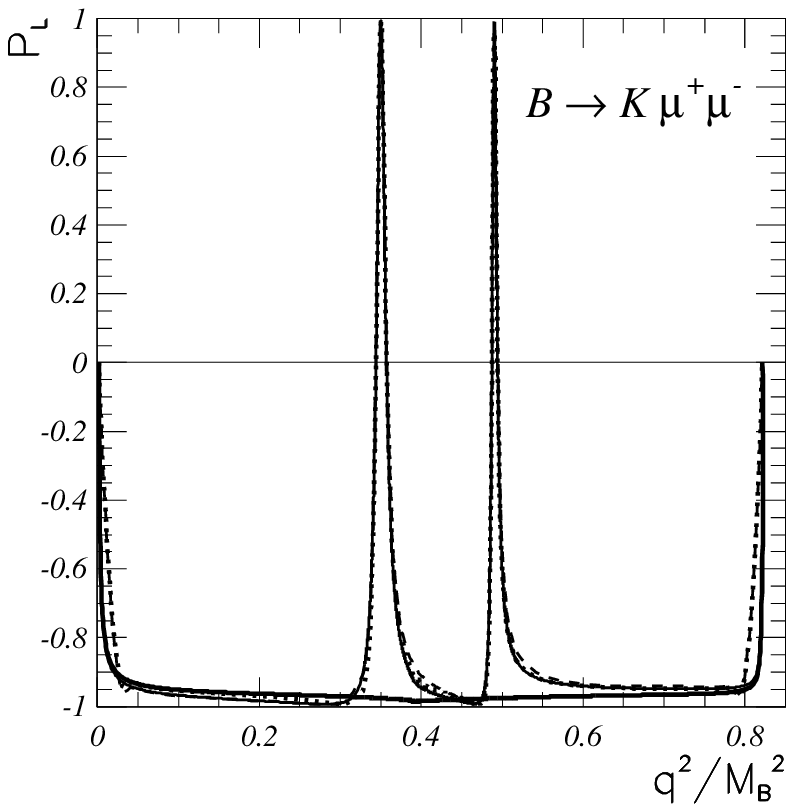,width=7.5cm}} 
\caption{\label{fig:afb-c9}
A sensitivity of the $A_{FB}$ in $B\to K^*\ell^+\ell^-$ (a), 
$P_L$ in $B\to K^*\ell^+\ell^-$ (b), and $P_L$ in $B\to K\ell^+\ell^-$ (c) 
to the uncertainty in $C^{eff}_{9V}$. The GI-OGE Set of the form factors 
\protect\cite{mns1} and $\Delta C_{7\gamma}^{c\bar c}=0$ are used. 
The curves correspond to various parametrizations of 
$C_{9V}^{eff}(q^2)$. Notations see Fig. 1.}
\end{figure}
\end{center}

\newpage

\begin{center}
\begin{figure}
\begin{tabular}{cc}
\mbox{\epsfig{file=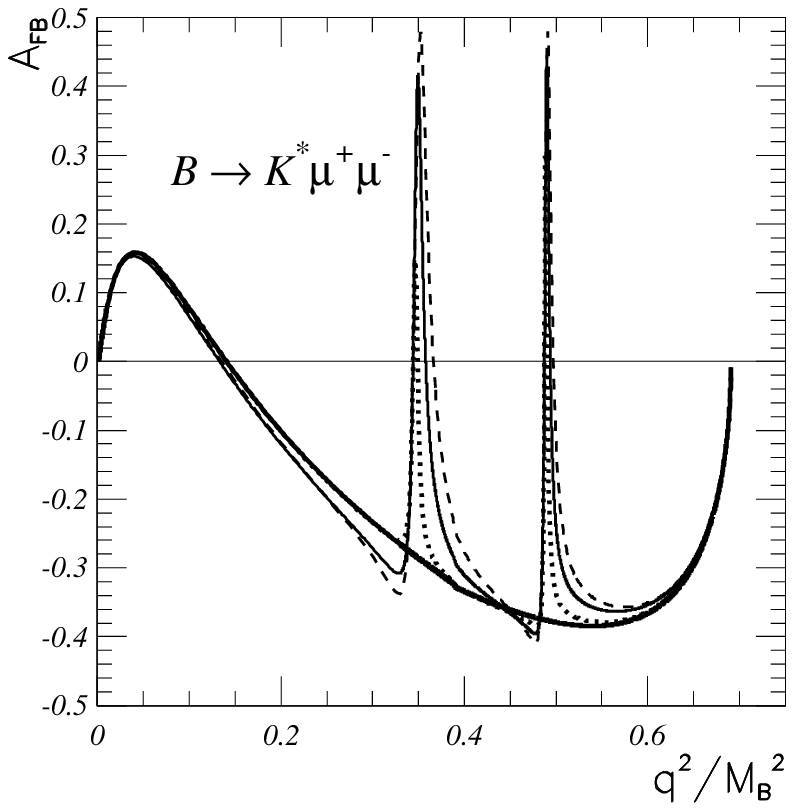,width=7.5cm}}
& 
\mbox{\epsfig{file=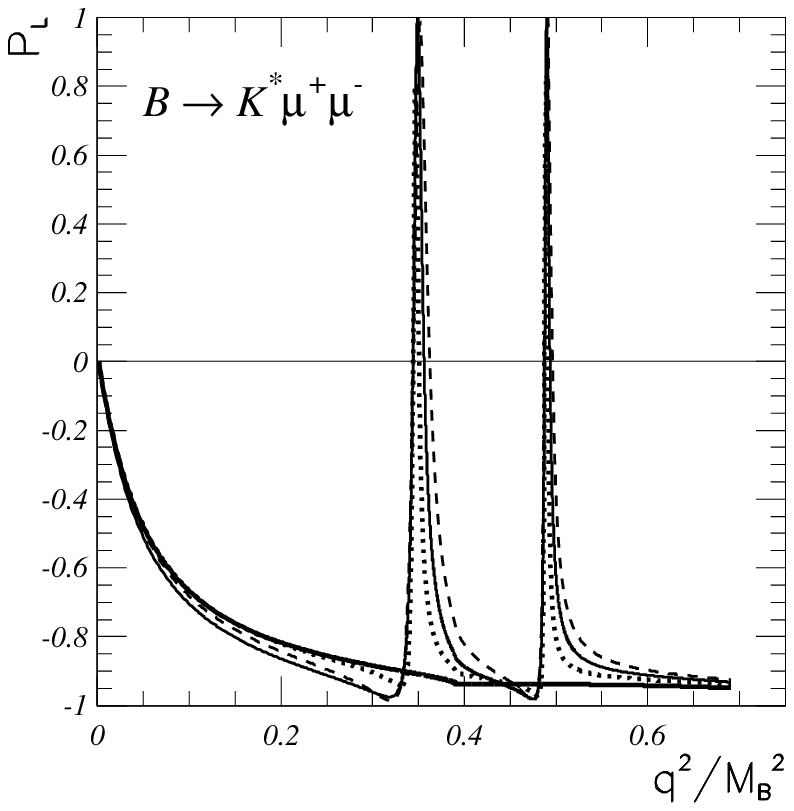,width=7.5cm}}
\end{tabular}

\mbox{\epsfig{file=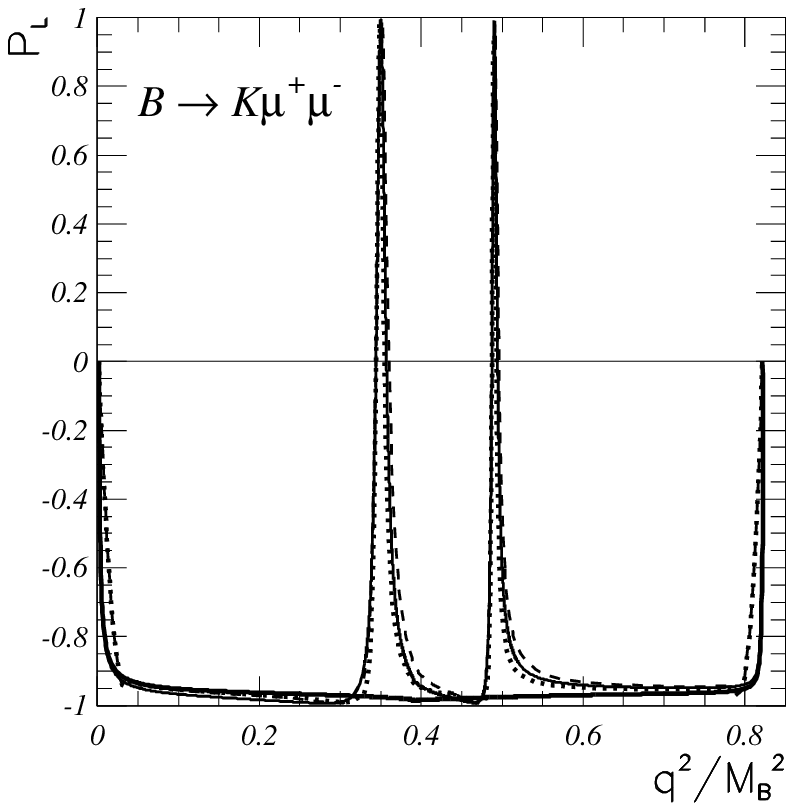,width=7.5cm}} 
\caption{\label{fig:afb-c7}
A sensitivity of the $A_{FB}$ in $B\to K^*\ell^+\ell^-$ (a), 
$P_L$ in $B\to K^*\ell^+\ell^-$ (b), and $P_L$ in $B\to K\ell^+\ell^-$ (c) 
to the nonfactorizable effects parametrized as additions to $C_{7\gamma}$: 
The $C^{eff}_{9V}$ from \protect\cite{ahmady} and the GI-OGE set of the form 
factors \protect\cite{mns1} are used. The curves correspond to 
$\Delta C_{7\gamma}^{c\bar c}=0.2 \Delta C_{9V}^{c\bar c}$ (dashed)
$\Delta C_{7\gamma}^{c\bar c}=0$ (solid)
$\Delta C_{7\gamma}^{c\bar c}=-0.2 \Delta C_{9V}^{c\bar c}$ (dotted). 
Bold line corresponds to $C_{9V}$ without the LD contributions and 
$\Delta C_{7\gamma}^{c\bar c}=0$. }
\end{figure}
\end{center}

\newpage

\begin{center}
\begin{figure}
\begin{tabular}{cc}
\mbox{\epsfig{file=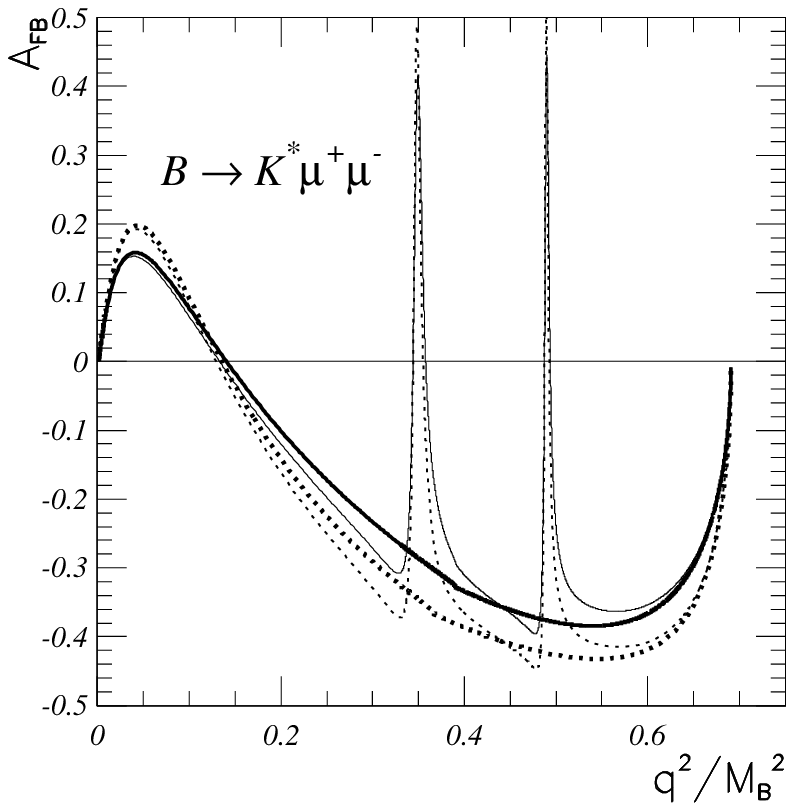,width=7.5cm}}
& 
\mbox{\epsfig{file=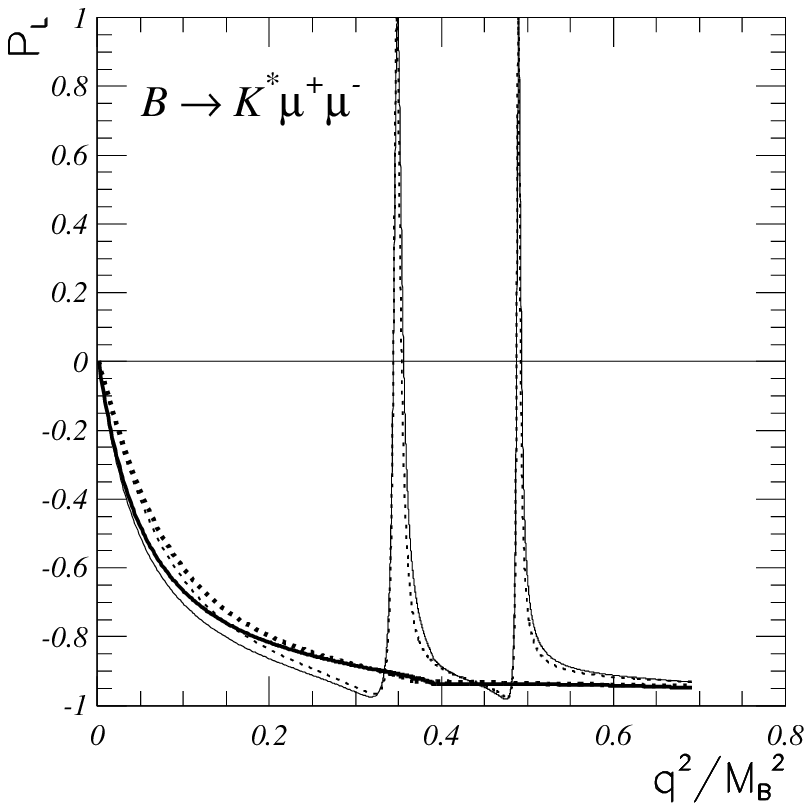,width=7.5cm}}
\end{tabular}
\caption{\label{fig:afb-ff}
A sensitivity of the $A_{FB}$ (a) and $P_L$ (b) in $B\to K^*\ell^+\ell^-$ 
to the meson transition form factors: solid - GI-OGE Set from Ref. 
\protect\cite{mns1}; 
dotted - lattice-constrained parametrizations from Ref. \protect\cite{lat}. 
The $C^{eff}_{9V}$ from \protect\cite{ahmady} and $\Delta C^{c\bar c}_{7\gamma}$ 
are used. 
Bold lines correspond to the $C_{9V}$ without LD contributions.} 
\end{figure}
\end{center}
\newpage
\begin{center}
\begin{figure}
\begin{tabular}{cc}
\mbox{\epsfig{file=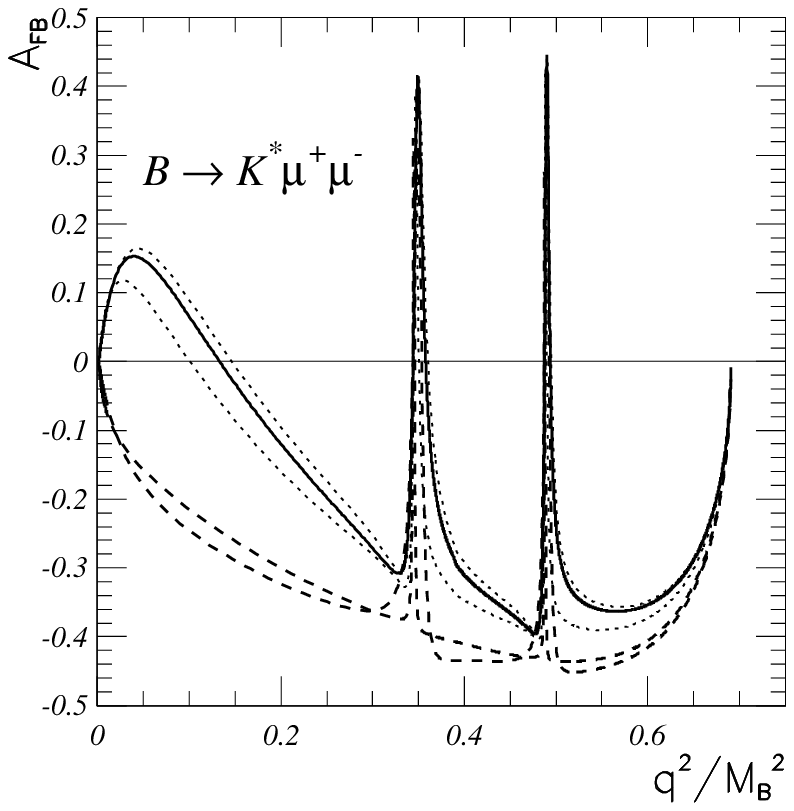,width=7.5cm}}
& 
\mbox{\epsfig{file=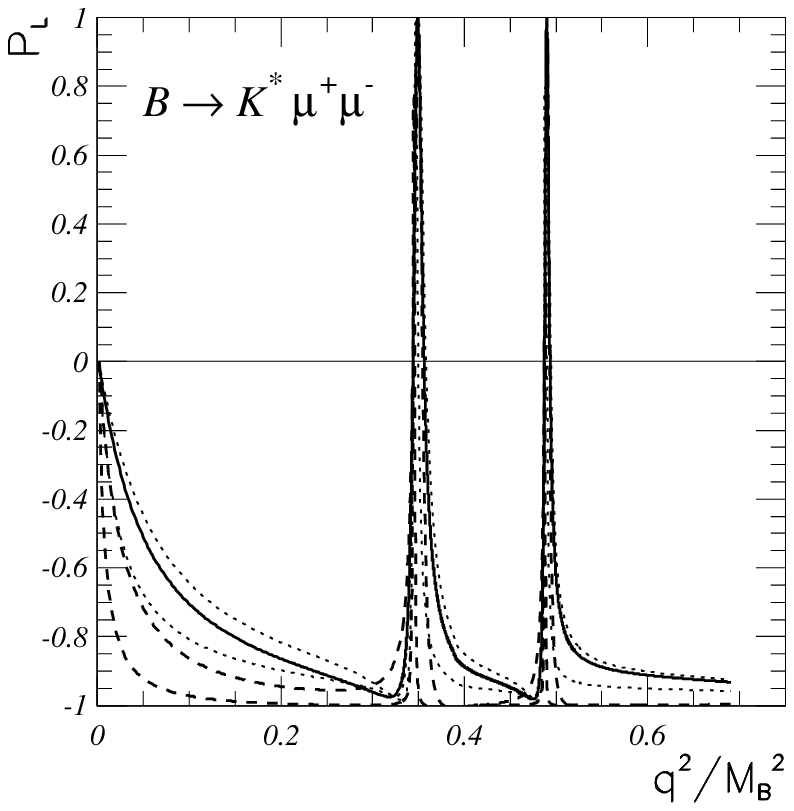,width=7.5cm}}
\end{tabular} 

\mbox{\epsfig{file=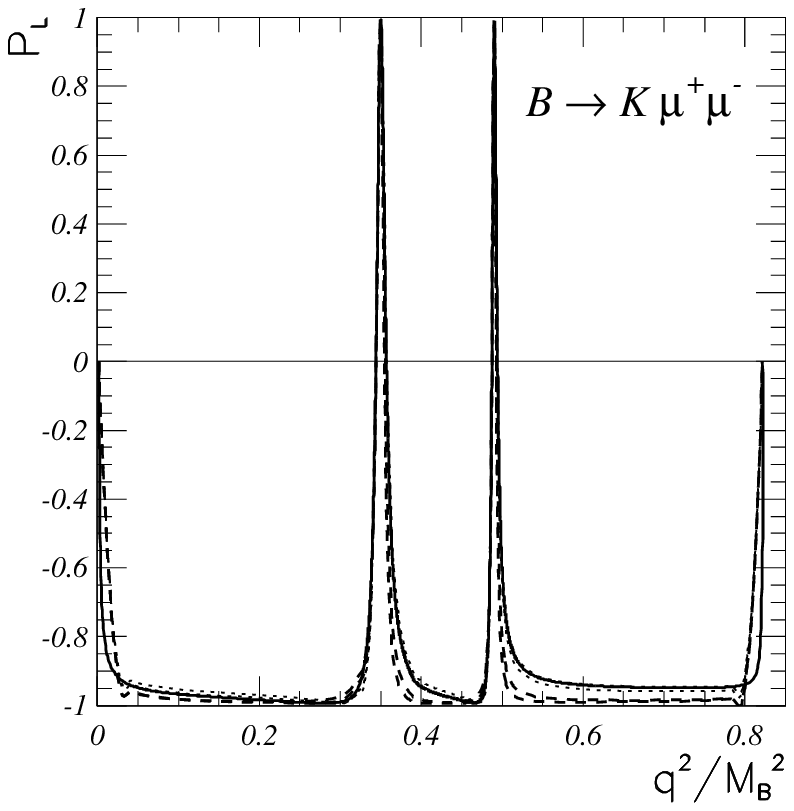,width=7.5cm}}
\caption{\label{fig:afb-mssm}
A sensitivity of the $A_{FB}$ (a) and $P_L$ (b) in $B\to K^*\ell^+\ell^-$ 
to the Wilson coefficient $C_{7\gamma}$: the asymmetries are evaluated  
for the GI-OGE Set of the form factors \protect\cite{mns1} and different values 
of 
$R_{7\gamma}=C_{7\gamma}(M_W)/C^{SM}_{7\gamma}(M_W)$ from the allowed region 
\protect\cite{misiak}: upper dotted line - $R_{7\gamma}=1.2$, 
lower dotted line - $R_{7\gamma}=0.4$, upper dashed line - $R_{7\gamma}=-2.4$, 
lower dashed line - $R_{7\gamma}=-4.2$, solid line - $R_{7\gamma}=1$ (SM). } 
\end{figure}
\end{center}
\end{document}